\documentclass[USenglish,twocolumn]{article}

\usepackage[utf8]{inputenc}
\usepackage[big]{dgruyter}
\usepackage{graphicx}
 
\begin{document}

\articletype{Research Article{\hfill}Open Access}

\author*[1]{Sergey A. Khaibrakhmanov}

\author[2]{\fbox{Alexander E. Dudorov}}

\affil[1]{Ural Federal University, 51 Lenina str, Ekaterinburg 620051, Russia, E-mail: khaibrakhmanov@csu.ru}

\affil[2]{Chelyabinsk State University, 129 Br Kashirinykh str, Chelyabinsk 454001}

\title{\huge Dynamics of magnetic flux tubes in accretion disks of Herbig Ae/Be stars}

\runningtitle{Magnetic flux tubes in disks of Herbig Ae/Be stars}

\begin{abstract}
{The dynamics of magnetic flux tubes (MFTs) in the accretion disk oftypical Herbig Ae/Be star with fossil large-scale magnetic field is modeled taking into account the buoyant and drag forces, radiative heat exchange with the surrounding gas, and the magnetic field of the disk. The structure of the disk is simulated using our magnetohydrodynamic (MHD) model, taking into account the heating of the surface layers of the disk with the  stellar radiation. The simulations show that MFTs periodically rise from the innermost region of the disk with speeds up to $10-12$~km~s$^{-1}$. MFTs experience decaying magnetic oscillations under the action of the external magnetic field near the disk's surface. The oscillation period increases with distance from the star and initial plasma beta of the MFT, ranging from several hours at $r=0.012$~au up to several months at $r=1$~au. The oscillations are characterized by pulsations of the MFT's characteristics including its temperature. We argue that the oscillations  can produce observed IR-variability of Herbig Ae/Be stars, which would be more intense than in the case of T~Tauri stars, since the disks of Herbig~Ae/Be stars are hotter, denser and have stronger magnetic field.
}
\end{abstract}

  \journalname{Open Astronomy}
\DOI{DOI}
  \startpage{1}
  \received{..}
  \revised{..}
  \accepted{..}

  \journalyear{2021}
  \journalvolume{..}
 
\maketitle
\section{Introduction}

Accretion disks are commonly observed around young stars. Analysis of contemporary observational data shows that accretion disks of young stars (ADYSs) evolve into protoplanetary disks (PPDs), in which conditions are favourable for planet formation. 

Polarization mapping of accretion disks and PPDs shows that they have large-scale magnetic field with complex geometry~\citep{li16}. Outflows and jets, which are ubiquitous in ADYSs, are indirect signs of the large-scale magnetic field in the system~\citep[see review by][]{frank14}. 
Robust measurements of the magnetic field strength in ADYSs are still not possible. There are indications that the magnetic field can be dynamically strong near the inner edge of the disk~\citep{donati05}. Analysis of the observational constraints on magnetic field strength from measurements of the remnant magnetization of meteorites~\citep{levi78} and Zeeman splitting of the CN lines~\citep{vlemmings19} shows that the magnetic field strength decreases with distance from the star. The observational data confirm predictions of the theory of fossil magnetic field, according to which the large-scale magnetic field of the accretion disks of young stars is the fossil field of the parent protostellar clouds~\citep[][]{dud95, fmft}.

MHD modeling of ADYSs have shown that strong toroidal magnetic field is generated in the innermost region of the ADYS, where thermal ionization operates and magnetic field is frozen in gas~\citep{fmfadys}. Runaway generation of the magnetic field in this region can be balanced by magnetic field buoyancy leading to the formation of magnetic flux tubes (MFTs), that float from the disk and carry away excess of its magnetic flux~\citep{kh17ppnl}. MFTs form in a process of magnetic buoyancy instability~\citep[also known as Parker instability, ][]{parker_book} in the stratified disk with strong planar magnetic field. Formation of MFT has been found both in MHD simulations of solar interior~\citep{vasil08} and simulations of the accretion disks~\citep{takasao18}.

Parker instability and rising MFTs can have different manifestations in the accretion disks~\citep[see review in][]{dud19mft}. \citet{kh17raa} and \cite{dud19mft} have shown that rising MFTs oscillate under certain conditions, and the oscillations can be the source of infrared (IR) variability of accretion disks of T~Tauri stars (TTSs). In this work, we further develop approach of Dudorov and Khaibrakhmanov and model the dynamics of the MFT in the accretion disk of typical Herbig Ae/Be star (HAeBeS).

Structure of the paper is following. In section~\ref{Sec:Model}, we outline the problem statement, describe our model of the dynamics of the MFT as well as the accretion disk model. In section~\ref{Sec:disk}, we present results of the simulations of the accretion disk structure. The structure of the disk of the HAeBeS is compared with those of the TTS. Section~\ref{sec:hydro} is devoted to the investigation of the dynamics of the MFT in absence of eternal magnetic field. Effect of the external magnetic field leading to magnetic oscillations of the MFT is investigated in section~\ref{sec:magneto}. We summarize and discuss our results in section~\ref{Sec:conc}.

\section{Model}
\label{Sec:Model}

\subsection{Problem statement}
We consider a toroidal MFT formed inside the accretion disk in the region of effective generation of the magnetic field. The dynamics of unit length MFT is modeled in the slender flux tube approximation. Cylindrical coordinates are adopted, $(r,\, 0,\, z)$, where $r$ is the radial distance from the center of the star, $z$ is the height above the midplane of the disk. The MFT is characterized by radius-vector $\mathbf{r}=(r,\, 0,\, z)$, velocity vector $\mathbf{v} = (0,\, 0,\, v)$, cross-section radius $a$, density $\rho$, temperature $T$, and internal magnetic field strength $B$. The disk has density $\rho_{\rm e}$, temperature $T_{\rm e}$, pressure $P_{\rm e}$ and magnetic field strength $B_{\rm e}$.
The MFT starts its motion at some radial distance $r$ from the star and a height $z_0$ above the disk's midplane, $z=0$. The MFT moves in the $z$-direction under the action of buoyant and drag forces.

\subsection{Main equations}
We follow \citet{dud19mft} and use the system of equations describing the MFT dynamics taking into account the buoyant force, turbulent and aerodynamic drag, radiative heat exchange with the external gas, magnetic pressure of the disk,
\begin{eqnarray}
	\frac{{\rm d}{\bf v}}{{\rm d}t} &=& \left(1 - \frac{\rho_{{\rm e}}}{\rho}\right)\mathbf{ g} + \mathbf{ f}_{{\rm d}},\label{Eq:motion}\\
	\frac{{\rm d}{\bf r}}{{\rm d}t} &=& \mathbf{ v},\label{Eq:velocity}\\
	M_{{\rm l}} &=& \rho\pi a^2,\label{Eq:mass}\\
	\Phi &=& \pi a^2B,\label{Eq:mflux}\\
	{\rm d}Q &=& {\rm d}U + P_{{\rm e}}{\rm d}V,\label{Eq:dQ}\\
	P + \frac{B^2}{8\pi} &=& P_{{\rm e}},\label{Eq:pbal2}\\
	\frac{{\rm d}P_{{\rm e}}}{{\rm d}z} &=& -\rho_{{\rm e}}g_z,\label{Eq:disc}\\
	U &=& \frac{P_{{\rm e}}}{\rho(\gamma - 1)} + \frac{B^2}{8\pi\rho}, \label{Eq:eos_kalor}
\end{eqnarray}
where $\mathbf{ f}_{{\rm d}}$ is the drag force, $M_{{\rm l}}=$~const is the  mass per unit length of the MFT, $\Phi=$~const is the magnetic flux of the MFT, $Q$ is the quantity of heat per unit mass of the MFT, $U$ is the energy of the MFT per unit mass, $g_z$ is the vertical component of stellar gravity, $\gamma$ is the adiabatic index.

Equations of motion (\ref{Eq:motion}, \ref{Eq:velocity}) determine dependences ${\bf v}(t)$ and ${\bf r}(t)$. Differential equations describing evolution of the MFT's density and temperature can be deduced by taking time derivative of the energy equation (\ref{Eq:dQ}) and pressure balance (\ref{Eq:pbal2}) and using the equation of the hydrostatic equilibrium of the disk (\ref{Eq:disc}). We define the rate of heat exchange as $h_{\rm c} = dQ/dt$ and estimate it in the diffusion approximation,
\begin{equation}
h_{{\rm c}} \simeq  -\frac{4}{3\kappa_{{\rm R}}\rho^2}\frac{\sigma_{{\rm R}}T^4 - \sigma_{{\rm R}}T_{{\rm e}}^4}{a^2}.
\end{equation}
where $\kappa_{\rm R}$ is the Rosseland mean opacity adopted from~\citet{semenov03}, $\sigma_{\rm R}$ is the Stefan-Boltzmann constant.

We introduce non-dimensional variables
\begin{eqnarray}
	u = v/v_{{\rm a}}, & \tilde{z} = z / H, & \tilde{T} = T / T_{{\rm m}}, \nonumber\\ 
	\tilde{\rho} = \rho / \rho_{{\rm m}}, & \tilde{t} = t / t_{{\rm A}}, & \tilde{h}_{{\rm c}} = h_{{\rm c}} / h_{{\rm m}},\nonumber\\
	 \tilde{a} = a / H, &	 \tilde{B} = B / B_{{\rm e}}, & \tilde{g} = g_z/f_{{\rm a}} , \nonumber\\
	\tilde{f}_{{\rm d}} = f_{{\rm d}} / f_{{\rm a}}, & \tilde{P} = P / (\rho_m v_{{\rm a}}^2), & 
\end{eqnarray}
where $v_{{\rm a}}$ is the Alfv{\'e}n speed, $t_{{\rm A}} = H / v_{{\rm a}}$ is the Alfv{\'e}n crossing time,  $h_{{\rm m}}={\varepsilon}_{{\rm m}}/t_{{\rm A}}$, ${\varepsilon}_{{\rm m}}$ is the energy density of magnetic field, $f_{{\rm a}} = v_{{\rm a}} / t_{{\rm A}}$. All scales are defined at the midplane of the disk. Then the final equations of the MFT dynamics can be written as (tilde signs are omitted)
\begin{eqnarray}
	\frac{{\rm d}u}{{\rm d}t} &=& \left(1 - \frac{\rho_{{\rm e}}}{\rho}\right)g + f_{{\rm d}},\label{Eq:u}\\
	\frac{{\rm d}z}{{\rm d}t} &=& u,\label{Eq:xi}\\
		\frac{{\rm d}T}{{\rm d}t} &=& \frac{2\left(\gamma - 1\right)}{\beta}\times \nonumber\\
	& & \frac{h_{{\rm c}}\left(\dfrac{\beta}{2}T + C_{{\rm m}}\rho\right) + \rho_e gu\left(\dfrac{C_{{\rm m}}}{2} - \dfrac{P_{{\rm e}}}{\rho}\right)}{\dfrac{3-\gamma}{2}C_{{\rm m}}\rho + \dfrac{\beta}{2}T + \left(\gamma - 1\right)\dfrac{P_{{\rm e}}}{\rho}},\label{Eq:tT}\\
	\frac{{\rm d}\rho}{{\rm d}t} &=& -\frac{\rho_{{\rm e}} gu + (\gamma - 1)h_{{\rm c}}\rho}{\dfrac{3-\gamma}{2}C_{{\rm m}}\rho + \dfrac{\beta}{2}T + \left(\gamma - 1\right)\dfrac{P_{{\rm e}}}{\rho}},\label{Eq:trho}\\
	a &=& C_{{\rm a}} \rho^{-1/2},\label{Eq:ta}\\
	B &=& C_{{\rm B}} \rho,\label{Eq:tB}
\end{eqnarray}
where $\beta$ is the midplane plasma beta, $C_{\rm m}=B_0^2/4\pi\rho_0^2$, $C_{\rm a} = \tilde{a}_0\tilde{\rho}_0^{1/2}$, $C_{\rm B} = \tilde{B}_0/\tilde{\rho}_0$.

Ordinary differential equations (\ref{Eq:u}--\ref{Eq:trho}) together with the algebraic equations (\ref{Eq:ta}, \ref{Eq:tB}) form closed system of equations describing the dynamics of the MFT. Equations (\ref{Eq:u}--\ref{Eq:trho}) are supplemented by the initial conditions $u(t=0)=0$, $z(t=0)=z_0$, $T(t=0)=T_{\rm e}$, $\rho(T=0)=\rho_0$, $a(t=0)=a_0$, $B(t=0)=B_0$. Values $z_0$ and $a_0$ are the free parameters of the model, while the initial density $\rho_0$ is calculated from the pressure balance (\ref{Eq:pbal2}) at $t=0$. Initial magnetic field strength $B_0$ is specified through the initial plasma beta inside the MFT, $\beta_0$, which is also a free parameter.

\subsection{Model of the disk}

The distributions of the density, temperature and magnetic field in the disk are calculated using our MHD model of the AD~\citep{fmfadys, kh17}. The disk is considered to be geometrically thin and optically thick with respect to its own radiation. The mass of the disk is small compared to the stellar mass $M$. Inner radius of the disk is equal to the radius of stellar magnetosphere. Outer radius of the disk is determined as the contact boundary with the external medium.

The model is the generalization of \citet{ss73} model. In addition to the solution of \citet{ss73} equations for the low-temperature opacities, we solve the induction equation for magnetic field taking into account Ohmic dissipation, magnetic ambipolar diffusion, magnetic buoyancy and the Hall effect. The ionization fraction is calculated following~\citet{dud87} taking into account thermal ionization, shock ionization by cosmic rays, X-rays and radionuclides, as well as radiative recombinations and recombinations onto dust grains. 

Vertical structure of the disk is determined from the solution of the hydrostatic equilibrium equation (\ref{Eq:disc}) for polytropic dependence of the gas pressure on density,
\begin{eqnarray}
	\rho_{{\rm e}}(z) &=& \rho_{{\rm m}}\left[1 - \left(\frac{z}{H_{\rm k}}\right)^2\right]^{\frac{1}{k-1}},\label{Eq:rhoe}\\
	T_{{\rm e}}(z) &=& T_{{\rm m}}\left[1 - \left(\frac{z}{H_{\rm k}}\right)^2\right],\label{Eq:Te}
\end{eqnarray}
where 
\begin{equation}
H_{\rm k} = \sqrt{\frac{2k}{k-1}}H,
\end{equation}
$\rho_{{\rm m}}=\rho_{{\rm e}}(z=0)$, $T_{{\rm m}}=T_{{\rm e}}(z=0)$ are the density and temperature in the midplane of the disk, $k=1+1/n$, $n$ is the polytropic index, scale height $H=v_{{\rm s}}/\Omega_{{\rm k}}$,
\begin{equation}
	\Omega_{{\rm k}} = \sqrt{\frac{GM_{\star}}{r^3}}
\end{equation}
is the Keplerian angular velocity.

We consider that there is an optically thin hydrostatic corona above the optically thick disk. The corona's temperature is determined by heating due to absorption of stellar radiation,
\begin{equation}
T_{\rm c} = 185 \left(\frac{f}{0.05}\frac{L}{1\,L_\odot}\right)^{1/4}\left(\frac{r}{1\,\rm{au}}\right)^{-1/2}\, \rm{K},
\end{equation}
where $f$ is the fraction of the stellar radiation flux intercepted by the disk, $L$ is the stellar luminosity~\citep[see][]{akimkin12}. Transition from the disk to corona is characterized by an exponential change in temperature over the local scale height $H$ in accordance with the results of detailed modeling of the vertical structure of the accretion disks~\citep[see][]{vorpav17}.

The model of the disk has two main parameters: turbulence parameter $\alpha$ and mass accretion rate $\dot{M}$. 

\subsection{Model parameters and solution method}

Ordinary differential equations (\ref{Eq:u}--\ref{Eq:trho}) of the model are solved with the Runge--Kutta scheme of the 4th order with step size control. 

Initially the MFT is in thermal equilibrium with external gas at $z_0=0.5\,H$. We performed a set of simulation runs for various initial radii of the MFT $a_0$, plasma beta $\beta_0$ and radial distances from the star $r$. Adopted ranges of the initial parameters are listed in Table~\ref{tab:params}. Adopted fiducial value of $r$ corresponds to the dust sublimation zone, where gas temperature is of $1500$~K.

\begin{table}
\centering
\caption{Model parameters: radial distance from the star, initial cross-section radius and plasma beta of the MFT. \label{Tab:1}}
\label{tab:params}
\begin{tabular}{ccc}
\hline 
quantity & range of values & fiducial value \\ 
(1) & (2) & (3) \\ 
\hline 
$r$ & $0.012-1$~au & $0.5$~au \\ 
$a_0$ & $0.01-0.4~H$ & $0.1\,H$ \\ 
$\beta_0$ & $0.01-10$ & $1$ \\ 
\hline 
\end{tabular} 
\end{table}

 We consider the accretion disk of Herbig Ae/Be star with mass $2\,M_\odot$, radius $1.67\,R_\odot$, luminosity $11.2\,L_\odot$, surface magnetic field strength $1$~kG, accretion rate $\dot{M}=10^{-7}\,M_\odot/\rm{yr}$, and turbulence parameter $\alpha = 0.01$. Adopted parameters correspond to the star MWC~480~\citep{db11, hub11}. Ionization and magnetic diffusivity parameters are adopted from the fiducial run in~\cite{kh17}.

\section{Results}
\label{Sec:Res}

\begin{figure*}[htb]
\begin{center}
\includegraphics[width=0.99\textwidth]{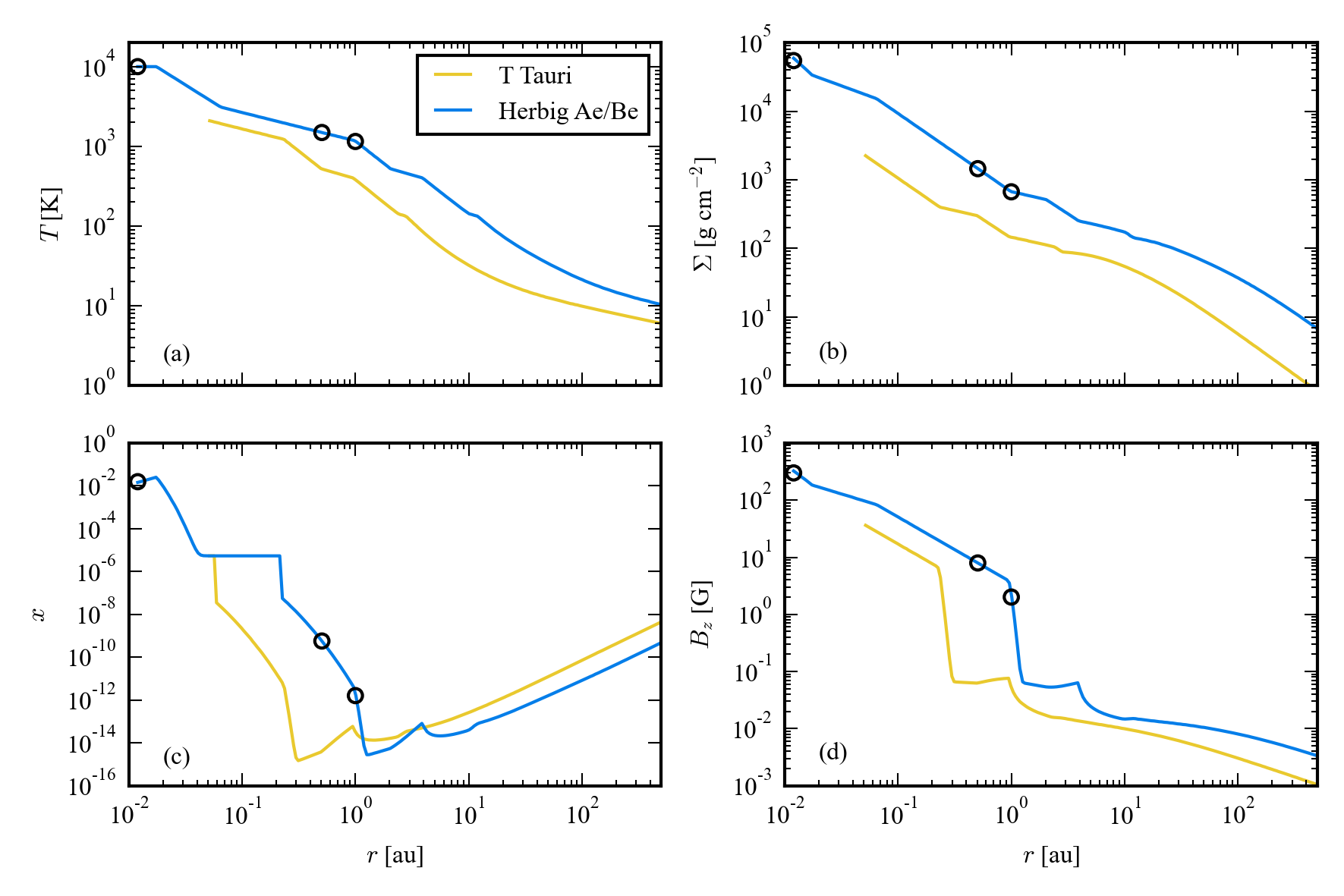}
\caption{Radial profiles of the midplane temperature (a), surface density (b), midplane ionization fraction (c) and midplane magnetic field strength (d) in the accretion disks of typical T Tauri star (yellow lines) and Herbig Ae/Be star  (blue lines). Empty circle markers show the points at which the modeling of the dynamics of the MFT was performed.}.
\label{fig:disk}
\end{center}
\end{figure*} 

\subsection{Radial structure of the disk}
\label{Sec:disk}
First of all, let us consider the structure of the accretion disk of HAeBeS in comparison with the structure of the disk of typical TTS according to our simulations.  Detailed discussion of the structure of TTS disks can found in our previous papers~\citep{fmfadys, kh17}.

In Figure~\ref{fig:disk}, we plot the radial profiles of midplane temperature $T_{\rm m}$, gas surface density $\Sigma$, midplane ionization fraction $x$ and magnetic field strength $B_z$. 

Figure~\ref{fig:disk} shows that the structures of the accretion disks of HAeBeS and TTS are qualitatively similar. 

Temperature and surface density are decreasing functions of distance, which can be represented as piece-wise power law profiles. The local slopes of the $T_{\rm m}(r)$ and $\Sigma (r)$ profiles are determined by the parameters of opacity dependence on gas density and temperature (see analysis of the analytical solution of model equations by~\citet{fmfadys}). 

The ionization fraction profiles $x(r)$ is non-monotonic and have minimum at $r_{\rm min}\approx 0.3$~au in the case of TTS and $1$~au in the case of HAeBeS. The ionization fraction is higher closer to the star, $r<r_{\rm min}$, due to thermal ionization of alkali metals and hydrogen. Growth of the $x$ further from the minimum, $r>r_{\rm min}$, is explained by decrease in gas density and corresponding increase in the intensity of ionizing radiation by external sources. Local peak in the $x(r)$ profiles at $r\approx 1$ (TTS) and $4$~au (HAeBeS) is due to evaporation of icy mantles of dust grains.

Intensity of the vertical component of the magnetic field $B_z$ generally decreases with distance. In the region of thermal ionization, $r<r_{\rm min}$, the magnetic field is frozen into gas and $B_z\propto \Sigma$. In the outer region, $r>r_{\rm min}$ magnetic ambipolar diffusion reduces magnetic field strength by 1--2 orders of magnitude as compared to the frozen-in magnetic field. For example, magnetic field strength is of $0.1$~G near the ionization minimum.

Comparison of the simulation results for HAeBeS and TTS shows that the accretion disk is hotter and denser in the former case at any given $r$. This is because the disk of HAeBeS has higher accretion rate, which leads to more intensive turbulent heating of the gas in the disk. As a consequence, the size of the innermost region, where runaway growth of the magnetic field is possible due to high ionization level, is more extended in the case of the HAeBeS. For adopted parameters, this region ranges from the inner boundary of the disk, $r_{\rm in}=0.012$~au, up to $r\approx r_{\rm min}=1$~au. Magnetic field strength is greater in the case of HAeBeS.

\subsection{MFT dynamics without external magnetic field}
\label{sec:hydro}
In this section we study the dynamics of the MFT in the disk of HAeBeS in absence of the magnetic field outside the MFT.

\begin{figure*}[htb!]
\begin{center}
\includegraphics[width=0.99\textwidth]{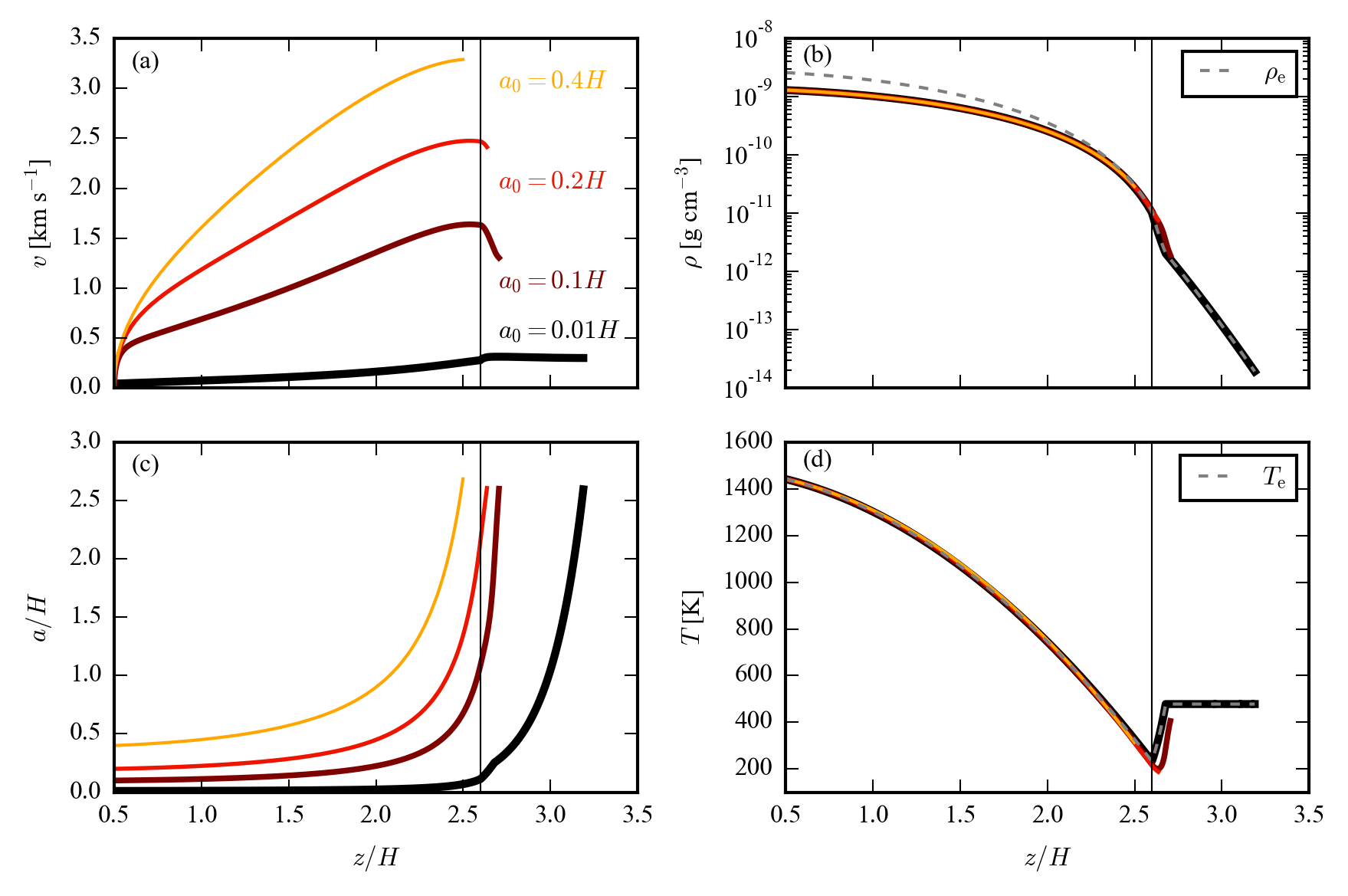}
\caption{Dynamics of the MFTs of various initial cross-section radii $a_0$ in absence of the external magnetic field. Dependences of the MFT's speed (panel a), density (b), cross-section radius (c) and temperature (d) on the $z$-coordinate are shown. Vertical lines show the surface of the disk. Grey dashed lines in panels (b) and (d) delineate corresponding profiles of the disk's density and temperature. Initial parameters of the MFT: $r=0.5$~au, $z_0=0.5\,H$, and $\beta_0=1$.}
\label{fig:dyn_vs_a0}
\end{center}
\end{figure*}

In Figure~\ref{fig:dyn_vs_a0}, we plot dependences of the MFT's speed, density, radius and temperature on the $z$-coordinate at $r=0.5$~au for different initial cross-section radii, $a_0=0.01$, $0.1$, $0.2$, and $0.4$~$H$.

Figure~\ref{fig:dyn_vs_a0}(a) shows that thinner MFT, $a_0=0.01$~H, is characterized by three stages of evolution. First the MFT accelerates inside the disk, then it rapidly decelerates near the surface, $z\approx 2.6\,H$, and after that it again rises with acceleration in the corona of the disk. Finally, the MFT dissipates in the corona, in a sense that its radius grows fast and becomes comparable with the half-thickness of the disk, as Figure~\ref{fig:dyn_vs_a0}(c) shows. Hence, MFTs will form outflowing magnetized corona of the disk, as in the case of TTS discussed by~\citet{dud19mft}. The MFT of intermediate initial radii, $a_0\sim 0.1-0.2\,H$, rise with higher speed, $v\approx 1-2$~km~s$^{-1}$, and dissipate right after rising to the corona without proceeding to the stage of further acceleration. Thick MFT with $a_0= 0.4\,H$ float with highest speeds up to $3$~km~s$^{-1}$ and dissipate near the surface of the disk.

Upward motion of the MFT is caused by the buoyancy force, which depends on the difference between internal and external densities, $\Delta \rho = \rho_{\rm e} - \rho$. Figure~\ref{fig:dyn_vs_a0}(b) shows that $\Delta \rho > 0$ and therefore the buoyant force is positive in all considered cases. The MFT expands and its density decreases during its motion in order to sustain the pressure balance. Near the surface of the disk and in the corona, $\Delta \rho$ approaches zero and therefore the MFT's speed decreases. Abrupt deceleration of the MFT after passing the surface of the disk, is caused by the abrupt disk's density drop in this region of transition from the disk to corona.

The MFT stays in thermal equilibrium, $T\approx T_{\rm e}$, during upward motion inside the disk, as Figure~\ref{fig:dyn_vs_a0}(d) shows. This is due to fast radiative heat exchange with the external gas. Departure form thermal equilibrium is observed only for the MFTs with $a_0=0.1-0.2\,H$ after their rising from the disk to the transtion region, where $T_{\rm e}$ grows up to the corona's temperature of $475$~K.

\begin{figure}
\begin{center}
\includegraphics[width=0.49\textwidth]{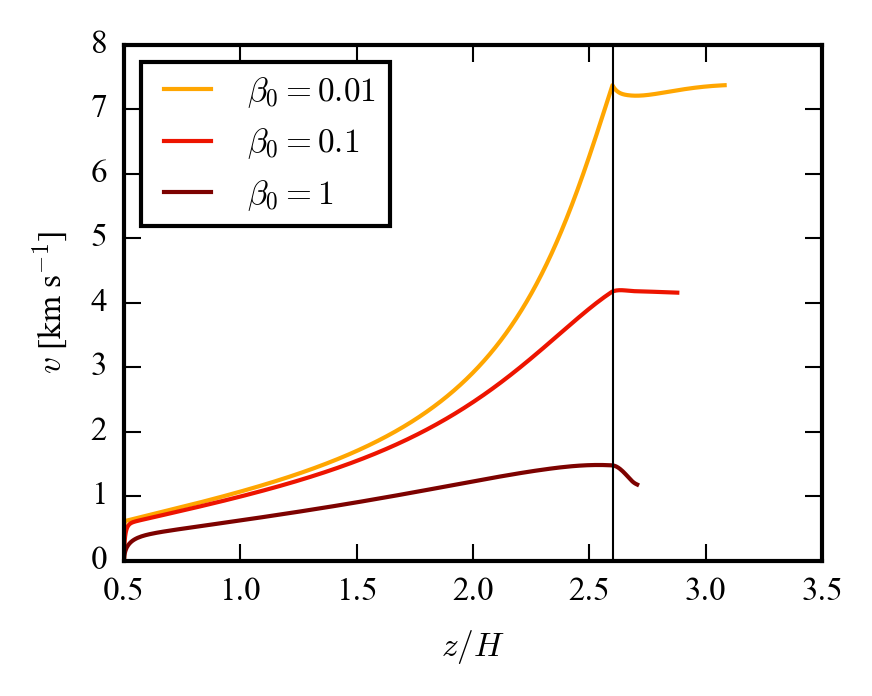}
\caption{Dependence of the MFT's speed on the $z$-coordinate for various initial plasma beta $\beta_0$ in runs without external magnetic field. Vertical line delineates the surface of the disk. Initial parameters: $r=0.5$~au, $z_0=0.5\,H$, and $a_0=0.1\,H$.}.
\label{fig:dyn_vs_beta0}
\end{center}
\end{figure}

In Figure~\ref{fig:dyn_vs_beta0}, we plot the dependence of the MFT's speed on the $z$-coordinate at $r=0.5$~au for $z_0=0.5\,H$, $a_0=0.1\,H$, and various initial plasma beta $\beta_0$. Figure~\ref{fig:dyn_vs_beta0} shows that the MFT with stronger magnetic field accelerate to greater speed. Maximum speed of $7-8$~km~s$^{-1}$ is achieved by the MFT with $\beta_0=0.01$.  The increase of the MFT's speed with $\beta_0$ is explained by the fact, that the more initial magnetic field strength of the MFT, the more initial $\Delta\rho$ and correspondingly the buoyant force is stronger. Closer to the star, $r<0.5$~au, the maximum speed is of $10-12$~km~s$^{-1}$, according to our simulations.

\subsection{Magnetic oscillations}
\label{sec:magneto}

In this section we investigate, how does the magnetic pressure outside the MFT influences its dynamics. In this case, external pressure $P_{\rm e}$ in (\ref{Eq:pbal2}) is a sum of gas pressure and magnetic pressure $B_{\rm e}^2/8\pi$. We assume that $B_{\rm e}$ is constant with $z$ and has magnitude of $B_z$.

\begin{figure*}[htb!]
\begin{center}
\includegraphics[width=0.99\textwidth]{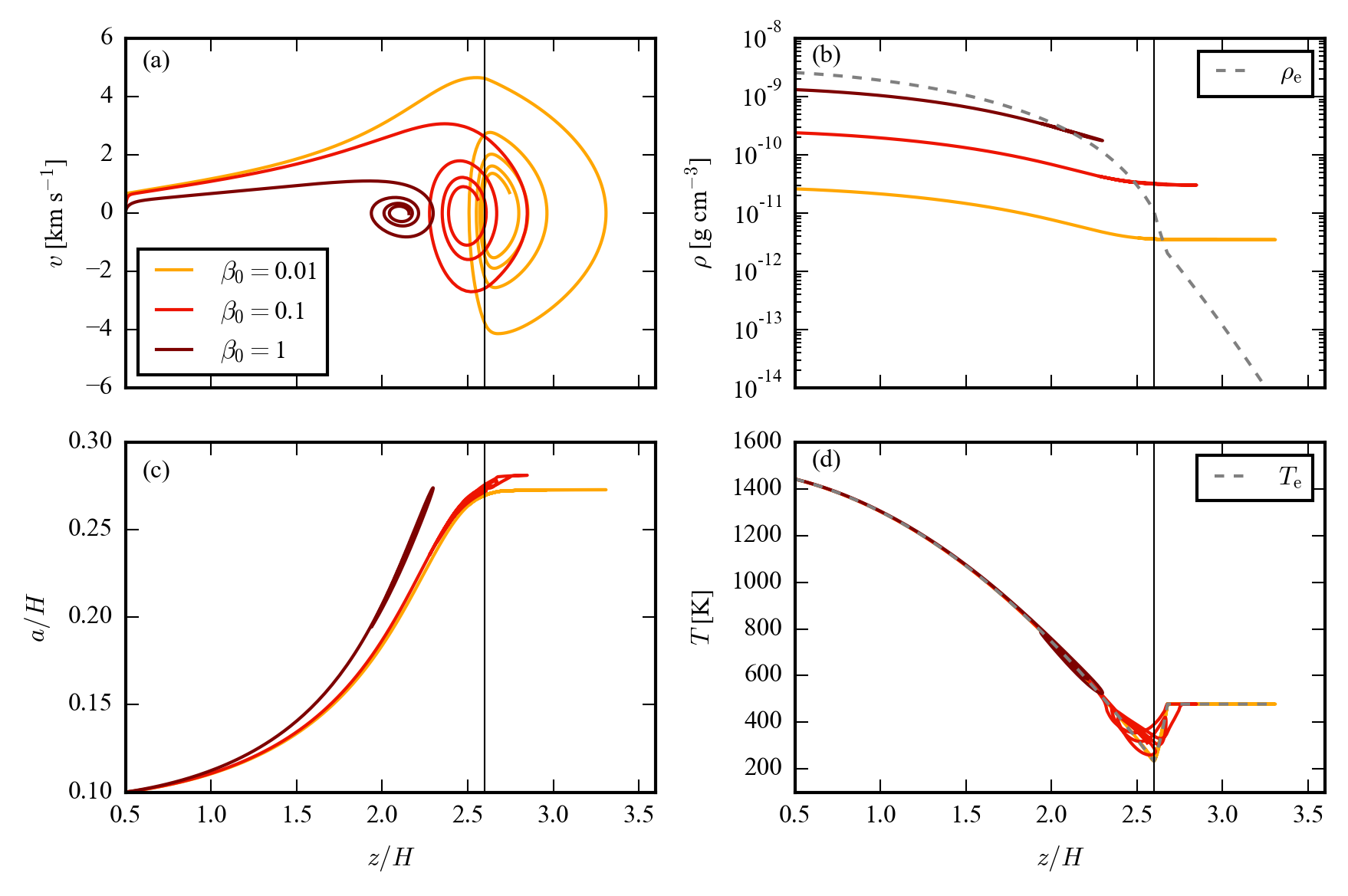}
\caption{Dynamics of MFTs with various initial plasma beta $\beta_0$ in presence of external magnetic field. Dependences of the MFT's speed (panel a), density (b), cross-section radius (c) and temperature (d) on the $z$-coordinate are plotted. Vertical lines show the surface of the disk. Grey dashed lines in panels (b) and (d) delineate corresponding profiles of the disk density and temperature. Initial parameters: $r=0.5$~au, $z_0=0.5\,H$, and $a_0=0.1\,H$.}
\label{fig:dyn_mf_vs_beta0}
\end{center}
\end{figure*}

In Figure~\ref{fig:dyn_mf_vs_beta0}, we present simulations results for the MFT at $r=0.5$~au with fiducial $z_0=0.5$~$H$, $a_0=0.1\,H$ and various initial plasma beta, $\beta_0$. The dependences of MFT's speed, density, radius and temperature on the $z$-coordinate are depicted. Magnetic field strength is of $8$~G at considered $r$.

Figure~\ref{fig:dyn_mf_vs_beta0} shows that the dynamics of the MFT  differs from the case with zero external magnetic field. The MFT floats with acceleration up to some height $z_{\rm o}$ near the surface of the disk, and then its motion becomes oscillatory: the MFT moves vertically up and down around the point $z_{\rm o}$. According to Figure~\ref{fig:dyn_mf_vs_beta0}(a), $z_{\rm o}\approx 2.2$, $2.5$, and $2.6$~$H$ for $\beta_0=1$, $0.1$, and $0.01$, respectively. Hence, the more $\beta_0$ the higher the point $z_{\rm o}$, around which the MFT oscillates. Magnitude of the MFT's speed decreases during oscillations as the MFT loses its kinetic energy due to the friction with the external gas.

When the MFT starts to oscillate, its expansion stops at some characteristic cross-section radius $a_{\rm o}$. In the considered case, this radius is of $0.3$~$H$, according to Figure~\ref{fig:dyn_mf_vs_beta0}(c). The radius of the MFT periodically increases and decreases with respect to  $a_{\rm o}$ during the oscillations, i.~e. the MFT pulsates. The magnitude of the radius variations decreases, i.~e. the pulsations decay with time. 

Figure~\ref{fig:dyn_mf_vs_beta0}(b) shows that the point $z_{\rm o}$ is a point of zero buoyancy, such that $\Delta \rho = \rho_{\rm e}-\rho>0$ at $z<z_{\rm o}$ and $\Delta\rho < 0$ at $z>z_{\rm o}$. This effect is caused by the contribution of the magnetic pressure outside the MFT to the overall pressure balance~(\ref{Eq:pbal2}). The external magnetic field $B_{\rm e}$ is constant with $z$, while the density of the disk $\rho_{\rm e}$ exponentially decreases. As a consequence, the magnetic pressure $B_{\rm e}^2/8\pi$ contribution to $P_{\rm e}$ also increases in comparison to the gas pressure. At $z\geq z_{\rm o}$, the external magnetic field becomes stronger than the magnetic field of the MFT, and consequently the MFT becomes heavier than the external gas. This result is similar to that found by~\citet{dud19mft} for the MFT in the accretion disks of TTS.

The beginning of the magnetic oscillations is characterized by violation of the thermal balance, $T\neq T_{\rm e}$, as Figure~\ref{fig:dyn_mf_vs_beta0}(d) shows. This means that the rate of radiative heat exchange is smaller than the rate of MFT's cooling due to adiabatic expansion. During the oscillations, the MFT's pulsations decay and radiative heat exchange ultimately equalizes the temperature $T$ and $T_{\rm e}$.

\begin{figure}
\begin{center}
\includegraphics[width=0.49\textwidth]{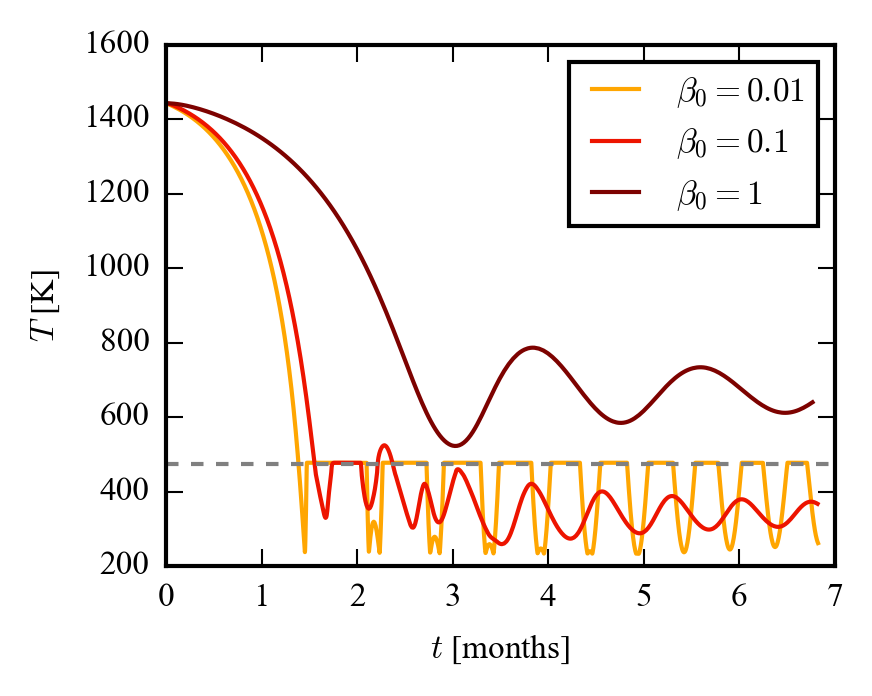}
\caption{Dependence of the MFT's temperature on time for various initial plasma beta $\beta_0$  in presence of external magnetic field. Horizontal dashed line delineates the temperature of the disk's corona. Initial parameters: $r=0.5$~au, $z_0=0.5\,H$, and $a_0=0.1\,H$.}.
\label{fig:dyn_mf_vs_beta0_t}
\end{center}
\end{figure}

In Figure~\ref{fig:dyn_mf_vs_beta0_t}, we plot corresponding  dependences of MFT's temperature on time. Figure~\ref{fig:dyn_mf_vs_beta0_t} clearly demonstrates the periodic changes in MFT's temperature during the magnetic oscillations. The period of oscillations increases with $\beta_0$ and lies in range from $0.5$~month for $\beta_0=0.01$ to $2$~months for $\beta_0=1$. 

Picture of the MFT's thermal evolution in the case $\beta_0=1$ can be described as simple decaying oscillations. In this case, the oscillations take place under the surface of the disk (see Figure~\ref{fig:dyn_mf_vs_beta0}(a)), and the MFT's temperature follows the  polytropic disk's temperature profile during its periodic upward and downward motion. The MFTs with $\beta_0\leq 0.1$ exhibit more complex behaviour characterized by non-harmonic oscillations of the temperature. In the case $\beta_0=0.01$, the maximum of each temperature pulsation is characterized by a constant $T=475$~K during time interval of $0.1-0.5$~months. Such a behaviour is explained by the fact, that the MFT's with $\beta_0\leq 0.1$ oscillate near the surface of the disk. The vertical profile of disk's temperature is non-monotonic in this region with minimum at $z_{\rm s}$, according to Figure~\ref{fig:dyn_mf_vs_beta0}(d). Therefore, the maximum $T$ in the oscillating MFT corresponds to the corona's temperature of $475$~K, while the minimum $T$ is achieved at some point below the surface of the disk.

\begin{figure*}[htb!]
\begin{center}
\includegraphics[width=0.99\textwidth]{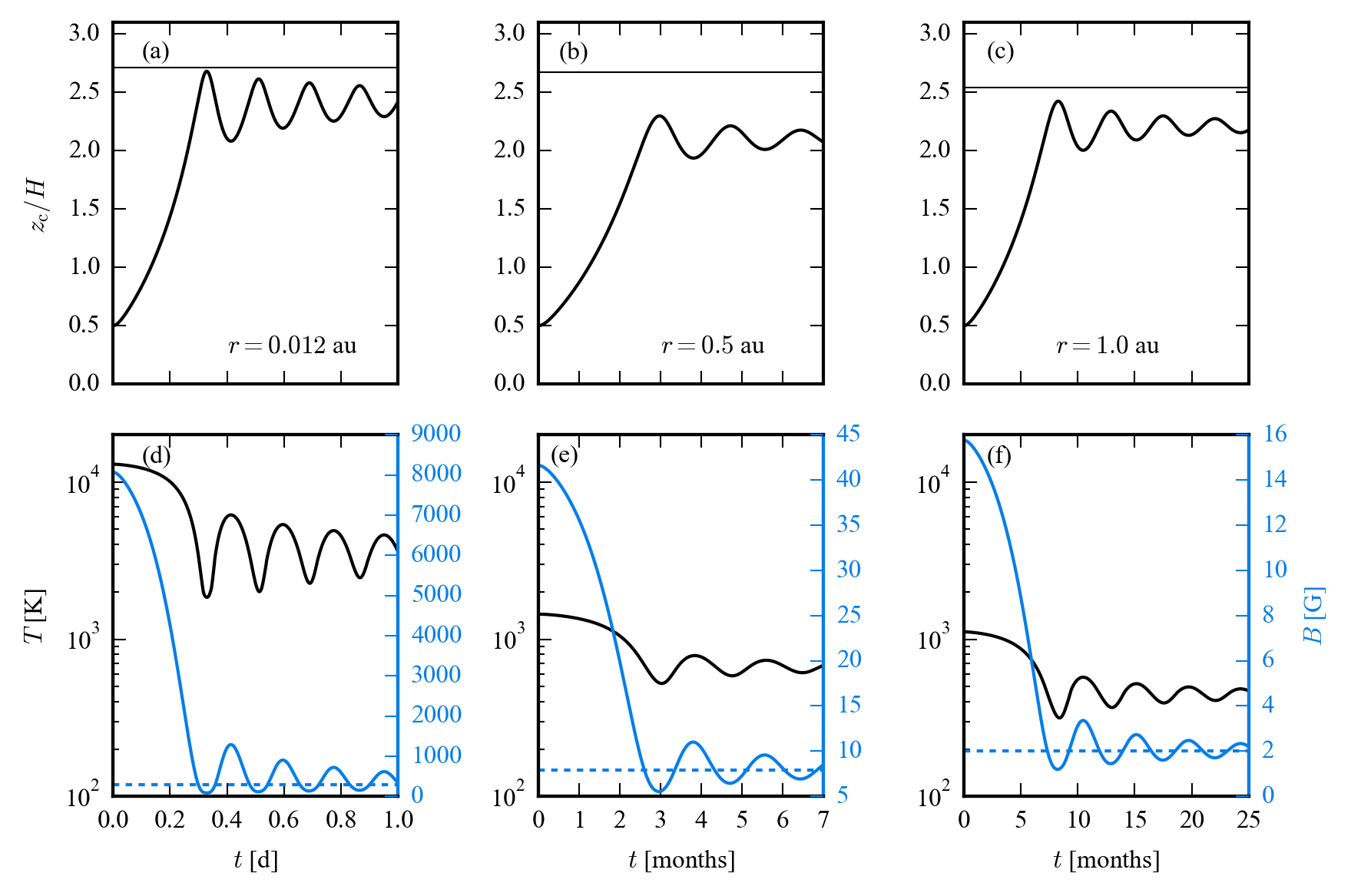}
\caption{Dynamics of the MFTs at different $r$ in presence of external magnetic field. Top row: the dependence of the $z$-coordinate of the MFT on time during its motion inside the disk at various radial distances $r=0.012$, $0.15$ and $1$~au (panels from left to right). Horizontal lines show the surface of the disk. Bottom row: corresponding dependences of temperature (left $y$-axis, black lines) and magnetic field strength (right $y$-axis, blue lines) on time. Horizontal dashed blue lines correspond to the external magnetic field $B_{\rm e}$. Initial radius and plasma beta of the MFT are $a_0=0.1\,H$ and $\beta_0=1$, respectively.}.
\label{fig:osc}
\end{center}
\end{figure*}

In order to investigate characteristic time scales of this process, we plot the dependence of the $z$-coordinate, temperature and magnetic field strength of the MFT on time in Figure~\ref{fig:osc}. The results for the MFT with fiducial parameters $z_0=0.5\,H$,  $a_0=0.1\,H$, and $\beta_0=1$ at various radial distances form the star are shown. Considered radial distances are marked with empty circles in $T(r)$, $\Sigma(r)$, $x(r)$, and $B(r)$ profiles in Figure~\ref{fig:disk}.

Figures~\ref{fig:osc}(a, b, c) show that the magnetic oscillations take place beneath the surface of the disk, at $z\sim 2-2.5\,H$ typically. The oscillations are found at every radial distance in the considered range, $r=0.012-1$~au. The period of oscillations $P_{\rm o}$ increases with $r$, such that $P_{\rm o}\approx 0.2$~d~$\approx 5$~hrs at $r=0.012$~au, $2$~months at $r=0.5$~au, and $5$ months at $r=1$~au. The amplitude of upward and downward motion decreases with time.

According to Figures~\ref{fig:osc}(d, e, f), the magnetic oscillations are accompanied by corresponding periodic changes in the MFT's temperature.The MFT heats up during the period of downward motion and cools down during its upward motion. These changes reflect the $z$-distribution of the disk's temperature $T_{\rm e}$, since the radiative heat exchange tends to keep the MFT in thermal balance with the external gas. The magnitude of the temperature fluctuations decreases with $r$. In  maximum, it ranges from $\Delta T\sim 3000$~K at $r=0.012$~au to $300$~K at $r=1$~au. 

Dependences $B(t)$ depicted in Figures~\ref{fig:osc}(d, e, f) show that the MFT's magnetic field strength decreases during its upward motion up to a point of the zero buoyancy. This decrease reflects the expansion of the MFT, $B\propto a^{-2}$ according to the magnetic flux conservations. During the following magnetic oscillations, the magnetic field strength periodically increases and decreases with respect to the value $B_{\rm e}$. This picture confirm above discussions that the magnetic oscillations arise due to the effect of external magnetic pressure near the point of zero buoyancy, which is characterized by the equality $B\approx B_{\rm e}$.

\section{Conclusions and discussion}
\label{Sec:conc}

We numerically modeled the dynamics of MFTs in the accretion disk of typical HAeBeS. The simulations were carried out in frame of the slender flux tube approximations using the model developed by~\citet{dud19mft}. This models allows to investigate the motion of the MFT in the direction perpendicular to the disk's plane taking into account the buoyant and drag forces, radiative heat exchange of the MFT with external gas, magnetic pressure of the disk.

The structure and characteristics of the accretion disk were calculated using our MHD model of the accretion disks~\citep[see][]{kh17}, which is based on the model of~\citet{ss73}. The vertical structure of the disk at each radial distance $r$ is calculated from the solution of the hydrostatic equilibrium equation for the case of polytropic gas. It is considered that the turbulent friction is the main heating mechanism inside the disk. There is optically thin corona above the optically thick disk. The temperature of the corona is determined by the heating of the gas with incident stellar radiation. We adopted that the fraction of the radiation flux intercepted by the disk's surface is constant, $f=0.05$, at every $r$. Transition from the disk to its corona is treated as hydrostatic region with exponential growth of gas temperature over the local scale height of the disk.

We adopted the parameters of the star and its accretion disk corresponding to the star MWC 480. This is a typical `isolated' HAeBeS, which was investigated in detail in different spectral ranges~\citep[see][]{sitko08, mendigutia13, tambov16, fernandez18}.

Our simulations have shown that the accretion disk of the HAeBeS is in general larger, denser and hotter than the accretion disk of typical TTS. This is because the disk in the former case is characterized by larger accretion rate. As a consequence, the magnetic field in the disk of HAeBeS is stronger than in the disk of TTS. The innermost region of the disk, where temperature is high enough for thermal ionization of alkali metals and hydrogen and where the magnetic field is frozen into gas, is more extended in the case of HAeBeS. This region ranges from $0.012$~au up to $r=1$~au in radial direction, under adopted parameters.

We modeled the dynamics of the MFT of various initial cross-section radii, $a_0$, and plasma beta, $\beta_0$, at several radial distances $r$ in the range $0.012-1$~au. The simulations have shown that the dynamics of the MFT in the accretion disk of the HAeBeS  is in general qualitatively similar to the case of typical TTS. In absence of the external magnetic field, MFTs rise from the disk with typical speeds up to $10-12$~km~s$^{-1}$ and form outflowing magnetized corona of the disk. Radiative heat exchange rapidly equalizes the temperatures inside and outside the MFT, $T\approx T_{\rm e}$, under all considered parameters. We did not found thermal oscillations of the MFT, caused by adiabaticity, unlike the case of the TTS, for which the thermal oscillations of the MFT with $a_0\sim 0.1$ and $\beta_0=1$ at $r<0.2$~au were found by~\citet{dud19mft}. Like in the case of TTS, MFTs transport excess of the disk's magnetic flux into its corona.

The pressure of the magnetic field outside the MFT halts upward motion of the MFT near the point, where internal and external magnetic fields are nearly equal. This point of zero buoyancy, $z_{\rm o}$, typically lies near the surface of the disk, $z_{\rm s}\sim 2.5-3$~$H$, where $H$ is the local isothermal scale height. The more initial magnetic field strength of the MFT, the higher the point $z_{\rm o}$ lies. After the MFT rises to this point, its motion becomes oscillatory. The MFT moves up and down around the point $z_{\rm o}$ and pulsates. The magnitude of these magnetic oscillations decreases with time because of the loss of the MFT's kinetic energy due to friction with external gas. The period of the oscillations increases with radial distance and ranges from few hours at the inner boundary of the disk, $r=0.012$~au, up to several months at $r=1$~au in the case of typical $a_0=0.1\,H$ and $\beta_0=1$. The oscillation period increases with $\beta_0$ at a given $r$.

Correspondingly, the temperature of the MFT experiences decaying oscillations around the value of local external temperature at the point $z_{\rm o}$. During the first few periods of the oscillations, temperatures inside and outside the MFT are not equal to each other, i.~e. radiative heat exchange is not efficient, and the MFT is practically adiabatic. But ultimately the heat exchange equalizes temperature of the MFT and external gas. The maximum magnitude of the temperature variations ranges from several thousand Kelvin at the inner edge of the disk to several hundred Kelvin at $r=1$~au. Temperature variations during each period of the oscillations may be non-harmonic and asymmetric, since the oscillations take place near the surface of the disk, where the dependence of the disk's temperature on the $z$-coordinate is complex and non-monotonic. 

Following original idea of~\citet{kh17raa}, we propose that the oscillations of MFTs can be a source of the emission variability as well as variable circumstellar extinction observed in young stars with accretion disks. Such a variability is a widespread feature of the accretion disks of TTS and HAeBeS~\cite[see][]{kospal12, flaherty16}, which also has been found for the MWC 480 star considered as a reference in our modeling. Generally speaking, periodically rising and oscillating MFTs could contribute to the variability of the emission in different spectral ranges emanating from the innermost region of the disk, where the magnetic field is frozen in into gas: $r=0.012-1$~au in the case of considered HAeBeS. The MFTs formed beyond the dust sublimation radius, $r=0.5$~au, could contain dust grains. In this case, temperature fluctuations of the oscillating MFT may cause the IR-variability of the disk. This assumption is supported by the observations of MWC 480 demonstrating the variations in the IR-flux  over $1-13\mu$m wavelength range~\citep{sitko08, fernandez18}. This radiation emanates from the dust sublimation zone and points to the changes of the disk's structure in this region. It should be noted, that inhomogeneities in
the disk centrifugal winds containing both gas and dust can cause
the variability of young stars' emission, in particular the variations in circumstellar extinction observed in young stars~\citep{tambov08}. Application of both models to specific systems with well-established variability is needed in order to determine relative role of various variability mechanisms.

In general, our results have shown that the accretion disks of HAeBeS have more extended region of the efficient generation of the magnetic field than in the case of TTS. The temperature of their corona is higher due to more intense stellar radiation. As a consequence, temperature variations in the oscillating MFTs has larger magnitude. Therefore, the IR-variability {caused by oscillating MFTs would be more intense in the case of accretion disks of HAeBeS as compared to TTS.

In order to investigate the connection between magnetic oscillations of MFTs and IR-variability of TTS and HAeBeS, we plan to calculate spectral energy distributions (SEDs) of the accretion disks taking into account variations of their structure due to the effect of rising MFTs. Interesting task is to model the synthetic light-curves of the accretion disks taking into account contribution of periodically rising MFTs into the IR flux of the disk.

\subsection*{Acknowledgements}
Authors thank anonymous referee for useful comments. The work is supported by the Russian Science Foundation (project 19-72-10012).

\end{document}